\documentclass{article}
\pdfoutput=1 

    \PassOptionsToPackage{numbers, compress}{natbib}
\usepackage[preprint]{neurips_2023}

\usepackage[utf8]{inputenc} % allow utf-8 input
\usepackage[T1]{fontenc}    % use 8-bit T1 fonts
\usepackage[hidelinks]{hyperref}       % hyperlinks
\usepackage{url}            % simple URL typesetting
\usepackage{booktabs}       % professional-quality tables
\usepackage{amsfonts}       % blackboard math symbols
\usepackage{nicefrac}       % compact symbols for 1/2, etc.
\usepackage{microtype}      % microtypography
\usepackage{xcolor}         % colors
\usepackage{graphicx} 
\usepackage{bm}
\usepackage{booktabs}
\usepackage{amsmath}
\usepackage{multirow}
\usepackage{bbm}
\usepackage{caption}
\usepackage{wrapfig}

\title{Linker-Tuning: Optimizing Continuous Prompts \\ for Heterodimeric Protein Prediction}
\author{
  Shuxian Zou \\
  MBZUAI \\
  \And
  Hui Li \\
  BioMap \\
  \And
  Shentong Mo \\
  MBZUAI \\
  \AND 
  Xingyi Cheng\\
  BioMap \\
  \And
  Eric Xing \\
  MBZUAI 
  \And 
  Le Song \\
  MBZUAI, BioMap \\
  \AND
  \texttt{\{shuxian.zou,le.song\}@mbzuai.ac.ae}
}

\begin{document}

\maketitle

\begin{abstract}

Predicting the structure of interacting chains is crucial for understanding biological systems and developing new drugs. Large-scale pre-trained Protein Language Models (PLMs), such as ESM2, have shown impressive abilities in extracting biologically meaningful representations for protein structure prediction. In this paper, we show that ESMFold, which has been successful in computing accurate atomic structures for single-chain proteins, can be adapted to predict the heterodimer structures in a lightweight manner.
We propose Linker-tuning, which learns a continuous prompt to connect the two chains in a dimer before running it as a single sequence in ESMFold. 
Experiment results show that our method successfully predicts 56.98\% interfaces on the i.i.d. heterodimer test set, with an absolute improvement of +12.79\% over the ESMFold-Linker baseline. Furthermore, our model can generalize well to the out-of-distribution (OOD) test set HeteroTest2 and two antibody test sets Fab and Fv while being $9\times$ faster than AF-Multimer.

%Notably, on the antibody heavy chain light chain (VH-VL) test set, our method successfully predicts all the heavy chain light chain docking interfaces, with 46/68 medium-quality and 22/68 high-quality predictions, while being $9\times$ faster than AF-Multimer.

\end{abstract}

\section{Introduction}
% biological background 
Proteins are large biomolecules essential to life. They are sequences compromised of 20 types of amino acids and fold into three-dimensional (3D) structures to carry out functions. Predicting the 3D structures of proteins from amino acid sequences is a long-standing challenge in computational biology. It is important for the mechanical understanding of protein functions as well as for designing new drugs. In 2021, AlphaFold2 (AF2) strikes a huge success in solving this challenge, achieving near experimental accuracy on protein structure prediction \cite{jumper2021highly}. However, this system heavily relies on Multiple Sequence Alignments (MSAs) to extract the evolutionary information, but MSAs are not always available or high-quality, especially for orphan proteins and fast-evolving antibodies \cite{lin2023evolutionary, wu2022high}. 

Inspired by the success of transformer language models in the field of Natural Language Processing (NLP), there is a line of work resorting to large-scale PLMs for protein structure prediction \cite{lin2023evolutionary, wu2022high, fang2022helixfold, chowdhury2022single}. These PLM-based models, such as ESMFold \cite{lin2023evolutionary}, take only amino acid sequences as input, eliminating the need for MSAs. Powered by PLMs, they show strong abilities in capturing protein structure information \cite{rao2020transformer, rives2021biological}. They are able to predict protein 3D structures at the atomic level with high accuracy while being an order of magnitude faster than AF2. However, these models are developed for predicting the structures of single-chain proteins and it is not clear how to use them to predict multi-chain protein structures.

To adapt these models for protein complex prediction, some researchers have proposed to use a poly-Glycine \textit{linker} to join chains and input the linked sequences to the model to predict complex structures \cite{ko2021can, tsaban2022harnessing}. The rationale is that the model should identify the linker segment as unstructured and fold the linked sequence in a similar way to multiple chains. Experimental result on AF2 shows that this approach is simple yet effective. However, for the PLM-based models, whether a linker is effective or not for protein complex prediction remains unexplored. In the work of ESMFold, they briefly mention that they use a 25-residue poly-Glycine linker (denoted as G25 in the following) to join different chains for a specific protein complex example \cite{lin2023evolutionary}. However, they do not test the performance of the linker systematically. Based on existing work, we would like to investigate the following questions in this paper:  1) \textit{How well can a G25 linker perform on protein complex prediction?} 2) \textit{Can we optimize the linker to achieve a better result? And how?}

% motivation 
Viewing proteins as the language of life, linkers in fact are the same things as prompts in natural language. Inspired by prompt engineering \cite{liu2023pre, liu2021gpt} in NLP, we propose Linker-tuning, which is to automatically learn a linker for the PLM-based model ESMFold on the task of heterodimeric protein structure prediction. Our goal is to find a linker that can link the two chains of a heterodimer so the structure prediction model can fold it similarly to a single-chain protein. How to best achieve this goal, however, is non-trivial and remains under-explored for the complicated protein structure prediction model. Through preliminary analysis, we find that it is better to place linker optimization at the Folding Module instead of at the PLM, which is different from intuition.

Considering ESMFold is a model with large-scale pre-trained PLM ESM2 that scales up to 15B parameters, to accelerate the linker learning procedure, we train and select our model on a proxy task called \textit{distogram prediction} \cite{senior2020improved}, a task that aims to predict inter-residue distance bins in the 3D space for each pair of residues in a given protein. After training, we test our learned linker on the 3D structure prediction task on three datasets to investigate the generalization ability of our method.

% contribution感觉写得不好
In summary, our main contributions are as follows: 
\begin{itemize}
    \item We propose Linker-tuning, to the best of our knowledge, the first prompt-tuning method for protein complex structure prediction. 
    \item We extend the natural amino acid linker to virtual space. We showcase that by just tuning the linker, our method outperforms the ESMFold-Linker baseline by large margins across all four test sets, including heterodimers with low sequence similarity and antibodies. 
    \item We find that our method is on par with AlphaFold-Linker on two general heterodimer test sets. Furthermore, it surpasses AlphaFold-Linker on two antibody test sets.
    \item We show that our method is 9$\times$ faster than AF-Multimer on the Fv test set although it is slightly worse than AF-Multimer. 
\end{itemize}

\section{Biological background} 

\paragraph{Linker} 
In biology, linkers are short amino acid sequences created in nature to separate multiple domains in a single protein \cite{reddy2013linkers}. Biologists have found that linkers rich in Glycine act as independent units and do not affect the function of the individual proteins to which they attach \cite{nagi1997inverse,deane2004tandem}. Therefore, we can use the Glycine-rich linker to join interacting chains to make it a single sequence, hoping it folds in the way they suppose to. Grounded in biological principles, we further extend the natural discrete linkers to virtual continuous linkers for better protein complex structure prediction.

\paragraph{Distogram and contact map}
The 3D structure of a protein is expressed as $(x, y, z)$ coordinates of the residues’ atoms in the form of a pdb file \cite{burley2018protein}. The distance between two residues in a protein 3D structure is defined as the Euclidean distance between their $C_{\beta}$ atoms ($C_{\alpha}$ for Glycine). Binning all the inter-residue distances in a protein into $k$ distance bins, we can obtain the distogram matrix \cite{senior2020improved}. For a protein with $L$ residues, the distogram $\bm{d}$ is an $L \times L$ matrix, with entry $\bm{d}_{ij}$ referring to the distance category of residue $i$ and $j$. In a coarser granularity, we can compute the contact map $\bm{c} \in \mathcal{R}^{L \times L}$, where $\bm{c}_{ij}=1$ means the distance between residue $i$ and $j$ is less than or equal to 8Å. For protein complexes, we are especially interested in the inter-chain contact maps where the contacts are formed by two residues from different chains. The inter-chain contact map reflects the interface of interacting proteins, which is essential for predicting the 3D structure of the complex.

%Particularly, for a protein complex with multiple chains, the inter-chain contacts are critical for us to understand how the single-chain proteins interact and fold in the 3D space.

\section{Related work}
% 篇幅控制在一页以内
\subsection{Protein structure prediction} 
\label{related work}
\paragraph{Single-chain protein structure prediction}
In recent years, single-chain protein structure prediction has attracted increasing attention from researchers in the Artificial Intelligence (AI) community, mainly due to the ground-breaking success of the deep learning model AF2. Deep learning based protein structure prediction methods can be classified into two main categories: 1) MSA-based methods, such as AF2, that take protein sequences and MSAs as input and predict 3D structures \cite{jumper2021highly, baek2021accurate, yang2020improved}; 2) PLM-based methods, such as ESMFold, that take only protein sequences as input and predict 3D structures \cite{lin2023evolutionary, wu2022high, fang2022helixfold, chowdhury2022single, wang2022single, wang2022xtrimoabfold, zhu2023uni, abanades2022immunebuilder, wang2023fast}. 
PLM-based methods do not rely on MSAs, which are time-consuming in searching homologs and not always available for some proteins like orphan proteins. Instead, they adopt large-scale pre-trained PLMs to learn evolutionary and structural meaningful representations for 3D structure prediction. In this work, we build our method upon PLM-based methods. Specifically, we adopt ESMFold \cite{lin2023evolutionary} as the backbone since its code and pre-trained weights are all released and convenient to use. The overall architecture of ESMFold contains two parts: 1) \textit{ESM2}: a PLM pre-trained with masked language modeling objective and scales up to 15B parameters; 2) \textit{Folding Module}: contains Folding Trunk (similar to Evoformer in AF2) and Structure Module (same as the one in AF2), which are responsible for structure folding.
%a transformer-based structure encoder with cubic complexity in sequence length, which is very much similar to Evoformer in AF2; and 3) \textit{Structure Module}: a structure decoder the same as the one in AF2. 

\paragraph{Multi-chain protein structure prediction}
%They play essential roles in most biological processes as proteins often function with interaction. 
In biology, multi-chain proteins are protein complexes formed by interacting single-chain proteins where the interactions are driven by the same physical forces as protein folding \cite{keskin2008principles}. In recent years, there is a line of work repurposing single-chain AF2 for protein complex structure prediction. The methods can be summarized into two main categories: 1) input-adapted methods that provide AF2 with pseudo-multimer inputs either by adding a large number to the residue\_index between chains to indicate chain break \cite{humphreys2021computed, bryant2022improved,  gao2022af2complex, mirdita2022colabfold} or using a linker to join chains \cite{ko2021can, tsaban2022harnessing}; and 2) training-adapted methods that retrain AF2 on multimeric proteins, such as AF-Multimer, the state-of-the-art (SOTA) method \cite{evans2022protein}. However, existing work mainly focuses on the MSA-based method AF2, with little attention being paid to the PLM-based methods. Recently, Zhu et al. (2023) proposes Uni-Fold MuSSe, which adapts ESM2 for multimer structure prediction \cite{zhu2023uni}. However, the model is fully fine-tuned, while our method only tunes a tiny number of extra parameters to the base model.

%On the one hand, the two types of methods either do not update any parameters, or update all parameters of the base model, while our method falls in between, adding only a tiny number of extra parameters to the base model. In this work, we focus on adapting the PLM-based methods for two-chain protein structure prediction, which has not yet been explored. 

\subsection{Prompt engineering}
In the NLP community, with the rise of large-scale pre-trained Language Models (LMs) such as GPT-3 \cite{brown2020language}, ``pre-train, prompt, and predict" has become a prevalent paradigm to steer the LM to perform a wide range of downstream tasks \cite{liu2023pre}. In this paradigm, the downstream tasks are reformulated in a form that is similar to the LM pre-training task using a textual prompt \cite{brown2020language, schick2021exploiting}. The key challenge in prompt-based learning is to find the right prompt for a specific task, termed ``prompt engineering". There is a line of work that automatically search the right prompts for downstream tasks \cite{shin2020autoprompt, gao2021making}. In particular, instead of  natural language prompts, some researchers propose to use continuous prompts, directly performing prompting in the embedding space of the LM \cite{li2021prefix, liu2021gpt}. In their experiment, continuous prompts achieve strong results in both language understanding and generation tasks. In this work, we follow the idea of continuous prompting, searching for the linkers in the continuous space.

%Since manual prompt search requires both domain expertise and an understanding of the inner workings of the LM, some researchers resort to automated prompt search, where our work falls in.

%Prompting does not need to update the LM weights, so it is parameter efficient. By reformulating the downstream task as a (masked) language modeling task, prompting closing the gap between inference and pre-training \cite{brown2020language, }. 

\section{Method: Linker-tuning}
\label{method}
To adapt the single-chain model for multi-chain protein structure prediction, we propose a lightweight adaptation method called Linker-tuning and a novel weighted distogram loss. The basic idea of our method is to optimize linkers, i.e., prompts, in the embedding space of ESMFold. 

%Linker-tuning, a lightweight adaptation method that only updates a tiny number of model parameters. 

%\Xingyi{Figure 1 looks somehow confusing, the token representations are not aligned it's Embedding Layer figure. It would be better to draw the parameters that need to be updated and place them in the middle so that r can be added by E(x).} 
%\Xingyi{If the implementation of past_key_value from subfix/prefix-tuning, how we encode position ids could be mentioned.}

\begin{figure}[t]
\centering
\includegraphics[width=1\columnwidth]{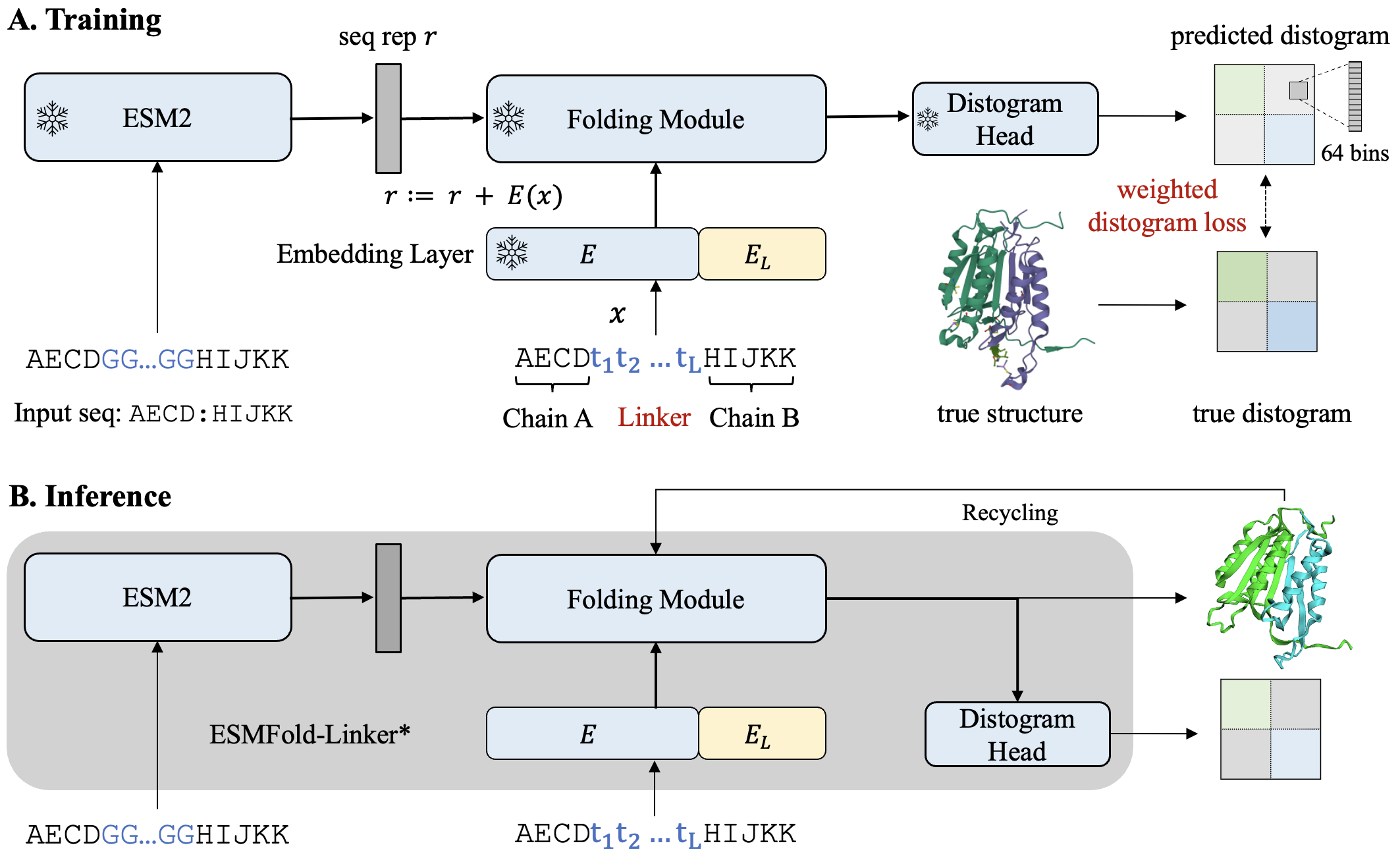}
\caption{\textbf{Overview of Linker-tuning method with ESMFold as backbone}. \textbf{(A)} Training. Based on ESMFold (shown in blue colors), we add a linker embedding module $E_L$ (shown in yellow colors) with linker length $L$. Given a protein with multiple chains, we add the linker specified in the linker embedding module between each chain before running it as a single chain through the ESMFold model. The model outputs a distogram with the linker part removed. We use a weighted distogram loss as the objective function to train the linker embedding module while freezing all the parameters in ESMFold. \textbf{(B)} Inference. After training, ESMFold with our linker embedding module can be treated as a whole black box model, denoted as ESMFold-Linker*. The input for this model is just protein sequences. And the model outputs a predicted distogram as well as all the atoms' 3D coordinates for the protein.}
\label{fig:method}
\end{figure}

\subsection{Problem formulation}
Continuous linker tuning of ESMFold for protein complex structure prediction is a continuous optimization problem. Our goal is to find a linker that maximizes the performance of ESMFold on protein complex prediction. To be specific, we first denote training data as $D_{train} = \{(x_1, y_1),...,(x_n, y_n)\}$ where $x_i =(x_i^A, x_i^B)$ and $x_i^A, x_i^B$ represent the amino acid sequences of two chains, $y_i$ is the structure of protein $x_i$. For a specified linker length $L$, the  linker optimization problem is defined as follows:
\begin{equation}
\bm{l}^* = \underset{\bm{l} \in E_L}{\arg \min} \, \frac1n \sum_{i=1}^n \mathcal{L}(x_i, y_i, \bm{l})
\end{equation}
where $\bm{l}$ denotes a linker, $E_L \subset \mathcal{R}^{L \times d}$ denotes a specific embedding space with embedding dimension of $d$, $\mathcal{L}(x_i, y_i, \bm{l})$ denotes complex structure prediction loss w.r.t. protein $(x_i, y_i)$ using linker $\bm{l}$. Therefore, the linker optimization is placed at the task level instead of at the instance level.
%\footnote{We use protein complexes with two chains (dimers) for simplicity in the description. Similarly, it can be applied to proteins with multiple chains.}

\subsection{Model architecture}
%\Xingyi{About "Too Deep for training". We can put it another way, for example, because the combined depth of two different representation modules~(Trunk and ESM2) with different structures actually exceeds 192 layers~(with cycle=3 ?), resulting in the gradient vanishing or exploding.}
Our method is implemented based on ESMFold, a PLM-based strong structure prediction model. As shown in Figure \ref{fig:method}, we place the continuous linker at Folding Module of ESMFold, which takes both the sequence representation from ESM2 and the amino acid sequence as input.
There are two main reasons that motivate us to place the continuous linker at Folding Module instead of at ESM2. First, we can utilize the pre-trained distogram head while avoiding backpropagating to the giant ESM2 model. If we put it on the ESM2 side, the combined depth of training will go up to 104 layers, making it easily suffer from gradient vanishing and exploding. Second, preliminary analysis on inter-chain contact prediction (shown in Table \ref{tab:hete2} and Appendix \ref{preliminary}) shows that using Folding Module on top of ESM2-3B increases prediction precision dramatically over ESM2-3B while ESM2-3B just performs slightly better ESM2-650M, implying that Folding Module is more sensitive to structure prediction and easier to control. 

We implement a plug-in linker embedding module, which contains $L \times d$ learnable parameters where $d$ is the embedding dimension of Folding Module. During training, only the linker embedding module is trainable, while all the original parameters in ESMFold are frozen. Therefore, ESM2 is just a sequence feature extractor that generates features for Folding Module. As shown in Figure \ref{fig:method}(A), we first use a poly-Glycine linker of the same length as the continuous linker to join different chains for the ESM2 input. Then we obtain the protein sequence representation and input it to Folding Module along with the chains connected by the continuous linker.
Finally, the distogram head outputs a probability distribution $\bm{p}_{ij}^D \in \mathcal{R}^{64}$ of each residue pair $(i, j)$ on 64 distance bins, which is used for computing the loss function. After training, we view ESMFold and the linker embedding module as a whole and name it as ESMFold-Linker*. As shown in Figure \ref{fig:method}(B), 
it can be used to predict the distograms as well as the 3D coordinates of all the residues for multi-chain protein sequences.

\subsection{Weighted distogram loss}
Intuitively, to predict the structure of a protein complex, we need to know two things: 1) the structures of each chain, on which ESMFold has been trained; and 2) the interaction interface between chains, which ESMFold has never seen before. Therefore, we propose to weight the intra-chain predictions and inter-chain predictions differently, with a focus on learning better interface between chains. 

Formally, let $N_A, N_B$ be the number of residues in two chains in a protein complex, $N = N_A+N_B$ be the total number of residues in the protein complex. Let $\bm{y}_{ij} \in \mathcal{R}^{64} $ denote the one-hot labels of the 3D space distance bins between residue pair $(i,j)$ and $\bm{p}_{ij} \in \mathcal{R}^{64} $ be the corresponding predicted probability. We define a weighted distogram loss for a protein complex as follows:
\begin{equation}
    \mathcal{L}(x, y, \bm{l}) = \mathcal{L}_1(x^A, y^A) + \mathcal{L}_1(x^B, y^B) + \lambda \mathcal{L}_2(x, y, \bm{l})
\end{equation}
where $\mathcal{L}_1(.,.)$ denotes the single-chain distogram loss given as follows:
\begin{equation}
    \mathcal{L}_1(x_A, y_A) = -\frac{2}{N_A(N_A+1)}\sum_{i=1}^{N_A}\sum_{j \geq i}^{N_A} \sum_{b=1}^{64} y_{ijb}log(p_{ijb}^D)
\end{equation}
and $\mathcal{L}_2(x,y,\bm{l})$ denotes the inter-chain distogram loss defined as follows:
\begin{equation}
    \mathcal{L}_2(x, y, \bm{l}) = -\frac{1}{N_A N_B}\sum_{i=1}^{N_A}\sum_{j=1}^{N_B} \sum_{b=1}^{64} y_{ijb}log(p_{ijb}^D)
\end{equation}
and $\lambda \geq 2$ is a hyperparameter controlling the attention we place on the interface of a protein complex. In our method, we use the weighted distogram loss as the training objective and validation metric.

% \subsection{\textcolor{blue}{Implementation details}}
% Due to the complexity of ESMFold architecture, it requires several optimizations to learn a linker better than the natural linker proposed by biologists. 

% Initialize the linker using the embedding of Glycine
% Tuning Initialized with Discrete Prompts

% Do not use binary protein-protein interaction. Do not use contact prediction. Do not use 3D structure prediction.

\section{Experiments}

\subsection{Experiment setting}

\paragraph{Datasets}
We mainly perform experiments on heteromers of two chains. For training, we use the dataset from APOC \cite{gao2013apoc}, which contains heterodimers released in the Protein Data Bank (PDB) before 2018-09-30. After filtering out similar sequences at a 40\% sequence identity threshold, it is split into train/valid/test\footnote{https://github.com/BioinfoMachineLearning/CDPred/tree/main/example/training\_datalists} sets by CDPred \cite{guo2022prediction}. We further filter out those proteins that contain missing $C_\beta$ coordinates ($C_\alpha$ for Glycine) in the pdb file. The resulting train/valid/test sample sizes are 2,946/193/172, respectively. The average number of residues in the test set is 367, with a maximum of 998. 
Furthermore, we use the largest blind test set HeteroTest2\footnote{https://zenodo.org/record/6647564\#.ZDWvMuxBxhE} from CDPred, which contains 55 heterodimers released in PDB between 2021-09-01 to 2021-10-20 \cite{guo2022prediction}. The average number of residues is 505, with a maximum of 979. 
In addition, we include an antibody antigen-binding fragments (Fab) test set, which is extracted from antibodies released in PDB between 2022-01-16 and 2022-04-13. It filters out samples that contain more than 10 consecutive missing residues in a chain. It also filters out chains whose length falls outside the range of [30, 1,024]. The average sequence length is 337 and the maximum is 454. We also leverage the antibody VH-VL test set from xTrimoDock \cite{luo2023xtrimodock}. It contains 68 samples released in PDB after 2022-02-01. Each sample consists of the variable domain of a heavy chain and the variable domain of a light chain, forming the Fragment variable (Fv) region. For convenience, we named it Fv. The average number of residues in Fv is 231, with a range of [223, 244].

\paragraph{Models}
% baseline: ESMFold-Linker, AlphaFold-Linker, HDOCK
% SOTA model: AF-multimer
% ESMFold: It is trained on single protein sequences released before 2020-05 
We use ESMFold-v1\footnote{https://dl.fbaipublicfiles.com/fair-esm/models/esmfold\_3B\_v1.pt} as our backbone model. ESMFold-v1 consists of a 3B ESM2 model and a 670M Folding Module, which is the largest yet publicly available ESMFold checkpoint. For the Linker-tuning method, the linker length $L$ is set to 25, equal to the length of the manual poly-Glycine linker. So the plug-in linker embedding module contains 0.027M parameters. We initialize the linker embedding using the embedding of Glycine. During training, only the linker embedding module is trainable, while all the original parameters in ESMFold are frozen. The hyperparameter $\lambda$ in the weighted distogram loss is set to 4. We train the model on a single Nvidia A100 80GB GPU with batch\_size=1 and num\_epoch=15. The protein sequences in the training set are cropped to 225 residues to fit in GPU memory using the multi-chain cropping algorithm from AF-Multimer \cite{evans2022protein}. The number of recycles is set to 1 during training to reduce computation. We use Adam optimizer with a learning rate of 5e-4. We select the best model based on the validation weighted distogram loss. During inference, the number of recycles is set to 3.
%In addition, the number of recycles is 1 and the residue index offset is 0 for the ESMFold model without otherwise specified. 

\paragraph{Baselines}
We compare our method with several baselines and one SOTA model as follows: 
\begin{itemize}
    \item ESMFold-Linker: ESMFold-v1 with chains joined by the G25 linker as input.
    %\item ESMFold-Gap: ESMFold-v1 with residue\_index\_offset set to 512.
    \item AlphaFold-Linker \cite{evans2022protein}: AF2 with a 21 residue repeated Glycine-Glycine-Serine linker.
    \item HDOCK \cite{yan2020hdock}: rigid docking with single chains predicted by AF2.
    \item AF-Multimer(v3 best) \cite{evans2022protein}: AF-Multimer contains five models that are trained on all protein structures released in PDB before 2021-09-30. We take the best prediction from the five AF-Multimer models.
\end{itemize}
%ESMFold-Linker-Gap: ESMFold-v1 with a 25 residue poly-Glycine linker and residue\_index\_offset set to 512.

\paragraph{Metrics}
For protein complex 3D structure prediction, we use DockQ \cite{basu2016dockq} to evaluate the quality of the predicted interfaces. As defined by Critical Assessment of PRediction Interactions (CAPRI), interfaces with $\text{DockQ}\geq 0.23$ means correct prediction. %interfaces with $0.23\leq \text{DockQ}<0.49$ means acceptable prediction, $0.49\leq \text{DockQ}<0.80$ means medium quality prediction, and $\text{DockQ} \geq 0.80$ means high-quality prediction. 
To evaluate the whole predicted protein complex structure rather than the interfaces, we adopt two commonly used global structure metrics, namely, Root Mean Squared Deviation (RMSD), and Template-Modeling Score (TM-Score) \cite{zhang2004scoring}. Besides, we use the top-$k$ precision as an evaluation metric for inter-chain contact prediction. We set $k=N_s/5$, where $N_s$ is the minimum chain length for a given protein complex.

% \begin{table}[t]
% \centering
% \caption{Overview of datasets.}
% \begin{tabular}{lcccc@{}}
% \toprule
% \multicolumn{1}{l}{Dataset} & \# chains & train & valid & test \\ \midrule
% Heterodimer \cite{gao2013apoc} & 2 & 2,946 & 193 & 172; 55 \\
% %Homodimer \cite{yan2021accurate} & 2 & 3,470 & 294 & 295 \\
% VH-VL~\cite{} & 2 & 7,481 & 300 & 171 \\ 
% AB-AG~\cite{luo2023xtrimodock} & 3 & 3,429 & 266 & 94 \\ 
% \bottomrule
% \end{tabular}
% \label{tab:data}
% \end{table}

\subsection{General heterodimer structure prediction}

Table \ref{tab:hete1} shows the PPI interface prediction quality of our methods and baselines on the heterodimer test set and HeteroTest2. On the i.i.d. heterodimer test set, ESMFold-Linker successfully predicts 76 out of 172 interfaces, reaching a success rate of 44.19\%. By optimizing the linker, our model ESMFold-Linker* successfully predicts 87 out of 172 interfaces, achieving a success rate of 50.58\%, with an absolute improvement of +6.39\% over ESMFold-Linker. By further incorporating a large residue gap, which adds a large number to the residue index in Folding Module, ESMFold-Linker*-Gap achieves a 56.98\% success rate in interface prediction, outperforming the baseline by +12.79\% absolute improvement. And it successfully predicts 5.24\% more interfaces than the MAS-based method AlphaFold-Linker. We further test our method on the OOD test set HeteroTest2 and observe consistent improvements in our methods over other baselines. These results demonstrate that our method is effective not only in predicting interfaces of i.i.d heterodimer dataset but also generalizes well to predicting interfaces of OOD heterodimer dataset. 

\begin{figure}[t!]
\centering
\includegraphics[width=1\columnwidth]{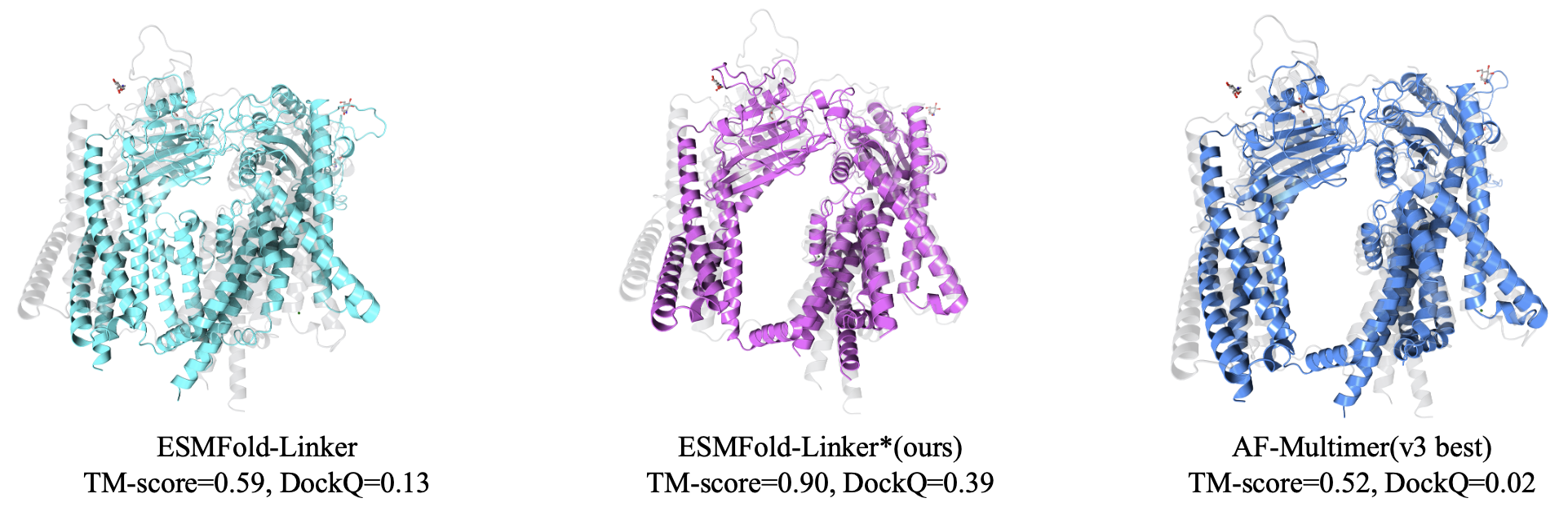}
\caption{Comparison of predicted structure quality of heterodimer \textbf{7D7F\_AD} by ESMFold-Linker, ESMFold-Linker*(ours), and AF-Multimer(v3 best). 7D7F is a membrane protein comprising 917 residues in the A and D chains. Structures are drawn using Protein Imager \cite{tomasello2020protein}. Gray indicates the ground truth structure. }
\label{fig:structure}
\end{figure}

Furthermore, we show the overall structure prediction results in Table \ref{tab:hete12}. On both the heterodimer test set and HeteroTest2, ESMFold-Linker*-Gap achieves the lowest RMSD among the linker-based methods. It also outperforms a strong docking method HDOCK in terms of both average DockQ score and RMSD. In addition, we compare it with the SOTA model AF-Multimer.\footnote{We use AF-Multimer v1 here because of the overlapping training data of AF-Multimer v3 and HeteroTest2. Since AF-Multimer v1 contains the heterodimer test set in its training data, we do not report the performance.} From Table \ref{tab:hete12}, we can see there is still a large gap between our method and the AF-Multimer(v1 best) on HeteroTest2. There are three main reasons responsible for this gap: 1) The base model for AF-Multimer is AF2, which is a model stronger than ESMFold(3B) in general, especially for those proteins that have high-quality MSAs; 2) AF-Multimer is a fully fine-tuned version of AF2 on a larger protein complex structure dataset while our model is a prompt tuning method trained only on the heterodimer dataset; 3) AF-Multimer ensembles five models, while we only use one model. However, our method can predict some proteins that are hard for both ESMFold-Linker and AF-Multimer. As shown in Figure \ref{fig:structure}, ESMFold-Linker* successfully predicts the interface 
of the membrane protein 7D7F\_AD with a DockQ score of 0.39 while ESMFold-Linker and AF-Multimer cannot predict the interface correctly.  

%On the i.i.d. heterodimer test set, ESMFold-Linker achieves a 0.316 DockQ score and a 10.762 r.m.s.d on average. By optimizing the linker, our model, i.e., ESMFold-Linker*, achieves a 0.359 DockQ score and a 9.189 r.m.s.d on average on the same test set, outperforming the ESMFold-Linker baseline by 13.61\% and 14.62\%, respectively. We further improve the ESMFold-Linker* by incorporating a large chain break, which adds a large number to the residue index in Folding Module. And the model ESMFold-Linker*-Gap achieves a 0.407 DockQ score and 8.594 r.m.s.d, outperforming ESMFold-Linker by 28.80\% and 20.14\%, respectively. On the OOD test set HeteroTest2, we observe similar results. ESMFold-Linker*-Gap outperforms ESMFold-Linker by 57.01\% DockQ score and 7.78\% r.m.s.d, respectively, suggesting that our learned linker can generalize well to OOD data.
%Compared to AlphaFold-Linker, a model that takes linked sequences and MSAs as input, our best model ESMFold-Linker*-Gap achieves similar DockQ scores on both test sets, with lower values in RMSD. Meanwhile, it outperforms the classic docking method HDOCK with AF2 predicted chains as input in terms of DockQ score and RMSD. 

\begin{table}[t]
\centering
\caption{PPI interface prediction quality on general \textbf{heterodimer} data.}
\label{tab:hete1}
\resizebox{1\columnwidth}{!}{
\begin{tabular}{@{}lcccccc@{}}
\toprule
 & \multicolumn{3}{c}{\textbf{Heterodimer test}} & \multicolumn{3}{c}{\textbf{HeteroTest2}} \\
\cmidrule(r){2-4}  \cmidrule(r){5-7}
 & \#Incorrect & \#Correct & Success rate & \#Incorrect & \#Correct & Success rate \\
\midrule
ESMFold-Linker & 96 & 76 & 44.19\% & 45 & 10 & 18.18\% \\
AlphaFold-Linker & 83 & 89 & 51.74\% & 44 & 11 & 20.00\% \\
ESMFold-Linker*(ours) & 85 & 87 & 50.58\% & 43 & 12 & 21.82\% \\
ESMFold-Linker*-Gap(ours) & 74 & 98 & \textbf{56.98\%} & 40 & 15 & \textbf{27.27\%} \\
\bottomrule
\end{tabular}}
\end{table}

\begin{table}[t]
\centering
\caption{Structure prediction results on general \textbf{heterodimer} data.}
\label{tab:hete12}
\resizebox{1\columnwidth}{!}{
\begin{tabular}{@{}lcccccc@{}}
\toprule
 & \multicolumn{3}{c}{\textbf{Heterodimer test}} & \multicolumn{3}{c}{\textbf{HeteroTest2}} \\
\cmidrule(r){2-4}  \cmidrule(r){5-7}
 & RMSD$\downarrow$ & TM-score$\uparrow$ & DockQ$\uparrow$ & RMSD$\downarrow$ & TM-score$\uparrow$ & DockQ$\uparrow$ \\
\midrule
ESMFold-Linker & 10.762 & 0.763 & 0.316 & 20.097 & 0.622 & 0.107 \\
AlphaFold-Linker & 9.381 & \textbf{0.827} & \textbf{0.418} & 20.954 & \textbf{0.707} & \textbf{0.170} \\
ESMFold-Linker*(ours) & 9.189 & 0.788 & 0.359 & 19.034 & 0.646 & 0.140 \\
ESMFold-Linker*-Gap(ours) & \textbf{8.594} & 0.795 & 0.407 & \textbf{18.534} & 0.651 & 0.168 \\
\midrule
HDOCK & 9.735 & 0.807 & 0.364 & 19.494 & 0.683 & 0.155 \\
AF-multimer(v1 best) & -- & -- & -- & \textbf{15.068} & \textbf{0.725} & \textbf{0.295}\\
\bottomrule
\end{tabular}}
\end{table}

\subsection{Antibody structure prediction}

\begin{table}[t]
\centering
\caption{Structure prediction results on \textbf{antibody} data.}
\label{tab:antibody}
\resizebox{1\columnwidth}{!}{
\begin{tabular}{@{}lcccccc@{}}
\toprule
 & \multicolumn{3}{c}{\textbf{Fab}} & \multicolumn{3}{c}{\textbf{Fv}} \\ 
\cmidrule(r){2-4}  \cmidrule(r){5-7}
 & RMSD$\downarrow$ & TM-score$\uparrow$ & DockQ$\uparrow$ & RMSD$\downarrow$ & TM-score$\uparrow$ & DockQ$\uparrow$ \\
\midrule
ESMFold-Linker & 5.027 & 0.851 & 0.546 & 1.459 & 0.955 & 0.737 \\
AlphaFold-Linker & 12.973 & 0.706 & 0.353 & 1.407 & 0.957 & 0.746 \\
ESMFold-Linker*(ours) & \textbf{2.948} & \textbf{0.901} & \textbf{0.649} & \textbf{1.388} & \textbf{0.959} & \textbf{0.753} \\
\midrule
HDOCK & 6.908 & 0.811 & 0.495 & 2.032 & 0.926 & 0.705 \\
xTrimoDock & -- & -- & -- & \textbf{1.264} & \textbf{0.965} & 0.775 \\
AF-Multimer (v3 best) & \textbf{2.752} & \textbf{0.908} & \textbf{0.697} & 1.287 & 0.963 & \textbf{0.779} \\ \bottomrule
\end{tabular}}
\end{table}

\begin{table}[t]
\centering
\caption{Structure prediction results on \textbf{Fv CDR loops}. RMSD is reported. }
\label{tab:cdr}
\resizebox{1\columnwidth}{!}{
\begin{tabular}{@{}lccccccc@{}}
\toprule
 & CDR-H1 & CDR-H2 & CDR-H3 & CDR-L1 & CDR-L2 & CDR-L3 & CDR-Mean \\ \midrule
ESMFold-Linker & 1.303 & 1.266 & 2.875 & 1.351 & 0.898 & 1.514 & 1.534\\
AlphaFold-Linker & \textbf{1.226} & 1.180 & 2.852 & \textbf{1.145} & 0.868 & \textbf{1.347} & 1.436\\
ESMFold-Linker*(ours) & 1.267 & \textbf{1.147} & \textbf{2.813} & 1.157 & \textbf{0.783} & 1.367 & \textbf{1.422} \\ 
\bottomrule
\end{tabular}}
\end{table}

Antibodies play a crucial role in the immune response by recognizing and binding to specific antigens. Given their importance, we further test how well our method generalizes to antibody structure predictions. Specifically, we test on two settings: 1) Fab structure prediction; and 2) Fv structure prediction.
Fab consists of two variable and two constant domains, with the two variable domains making up the variable fragment (Fv), which provides the antigen specificity of the antibody. 

Table \ref{tab:antibody} shows the structure prediction results of our methods and baseline methods on the Fab and Fv test sets. 
On the Fab test set, ESMFold-Linker achieves an average DockQ score of 0.546, outperforming AlphaFold-Linker and HDOCK by large margins. Equipped with the optimized linker, ESMFold-Linker* achieves an average DockQ score of 0.649, outperforming the ESMFold-Linker baseline by +18.86\%. Meanwhile, it achieves an average TM-score of 0.901, surpassing ESMFold-Linker by +5.88\%. It also reduces almost half of the RMSD compared to ESMFold-Linker. Besides, it successfully predicts 164 out of 171 interfaces, the same as AF-Multimer(v3 best). 
On the Fv test set, we observe similar results. Since most sequence variations associated with antibodies are found in the Complementarity-Determining Regions (CDRs), we further compute RMSD in the CDR loops on the Fv test set. As shown in Table \ref{tab:cdr}, our method achieves 1.422 mean r.m.s.d, with each CDR region consistently better than ESMFold-Linker. When compared to AlphaFold-Linker, our method achieves lower RMSD on CDR-H2, CDR-H3, and CDR-L2. These results demonstrate that our method trained on the general heterodimer dataset, generalizes well to antibody structure prediction.

\section{Analysis and discussion}

\paragraph{ESMFold-Linker* is 9$\times$ faster than AF-Multimer in inference}

We evaluate the speed of the model by testing on the Fv test set on a single Nvidia A100 80G GPU. As shown in Table \ref{tab:time}, ESMFold-Linker* makes a prediction on a protein with 231 residues in 3.5 seconds, $9\times$ faster than a single AF-Multimer model. In addition, the search process in CPU for constructing MSAs for a protein can take >10 min with the high-sensitivity protocols used by the published version of AF2, which further increases the time needed for AF-Multimer to predict protein structures. In contrast, ESMFold-Linker* does not require MSAs, which is convenient and fast.

\begin{table}[h]
\centering
\caption{Time analysis on the Fv test set.}
\begin{tabular}{@{}lccc@{}}
\toprule
(Seconds   per sample) & MSA search time & Model inference  time & Total time \\
\midrule
ESMFold-Gap & 0 & 2.7 & 2.7 \\
ESMFold-Linker & 0 & 3.5 & 3.5 \\
ESMFold-Linker*(ours) & 0 & 3.5 & 3.5 \\
\midrule
AF-Multimer(1 model) & >600 & 32.0 & >632.0 \\
\bottomrule
\end{tabular}
\label{tab:time}
\end{table}

\paragraph{Large chain break or linker, or both?}
%In the literature of input-adapted methods of AF2 as described in Section \ref{related work}, there are two major operations: 1) adding a large chain break to the residue index \cite{humphreys2021computed, bryant2022improved,  gao2022af2complex, mirdita2022colabfold}, and 2) using a linker to join each chain \cite{ko2021can, tsaban2022harnessing}. 
We perform an ablation study on ESMFold with chain break and linker to better understand the contribution of each operation. Table \ref{tab:hete2} shows the comparison of inter-chain contact prediction precision of ESMFold-based methods on the heterodimer test set and HeteroTest2.\footnote{The contact map probabilities are obtained from the predicted distogram probabilities by summing the probability mass in each distribution below 8.25{\AA}.} 
%We use CDPred \ref{tab:hete2} as a baseline. It is a strong inter-chain contact prediction model that takes sequences and ground-truth single-chain structures as input. 
As shown in Table \ref{tab:hete2}, it is hard to tell whether ESMFold-Linker or ESMFold-Gap is better. However, combining the two (ESMFold-Linker-Gap) provides significant performance gains over using either operation alone on both datasets. We observe similar effects in our method when incorporating chain break with the optimized linker. Compared to using a chain break, the major limitation of using a linker is that it increases the computation cost (shown in Table \ref{tab:time}). But we can enjoy the advantage of a large degree of freedom for improvement and better performance. Empirically, combining the two gives a better performance than just using each of them.

\begin{table}[t]
\centering
\caption{Comparison of inter-chain contact prediction results on \textbf{Heterodimer} data.}
\label{tab:hete2}
\resizebox{0.9\columnwidth}{!}{
\begin{tabular}{@{}lcccccc@{}}
\toprule
 \multicolumn{1}{l}{\multirow{2}{*}{(\%)}} & \multicolumn{3}{c}{Heterodimer test} & \multicolumn{3}{c}{HeteroTest2} \\
\cmidrule(r){2-4}  \cmidrule(r){5-7} 
 & top Ns/5 & top Ns/2 & top Ns & top Ns/5 & top Ns/2 & top Ns \\
\midrule
CDPred \cite{guo2022prediction} & -- & --  & -- & 22.87 & 20.17 & \textbf{17.51} \\
ESM2-650M-Linker & 12.02 & 9.89 & 8.33 & -- & -- & -- \\
ESM2-3B-Linker & 12.14 & 10.86 & 8.89 & -- & -- & -- \\
\midrule
ESMFold-Linker & 49.88 & 47.04 & 40.64 & 23.00 & 18.92 & 13.72 \\
ESMFold-Gap & 51.15 & 48.13 & 40.82 & 22.09 & 18.21 & 13.08 \\
ESMFold-Linker*(ours) & 57.55 & 53.04 & 44.37 & 27.11 & 22.14 & 15.46 \\
ESMFold-Linker-Gap & 57.72 & 53.44 & 45.41 & 25.20 & 19.84 & 14.66 \\
ESMFold-Linker*-Gap(ours) & \textbf{60.40} & \textbf{56.27} & \textbf{48.00} & \textbf{28.00} & \textbf{23.69} & 17.26 \\
\bottomrule
\end{tabular}}
\end{table}

\begin{figure}[t]
\centering
\includegraphics[width=1\columnwidth]{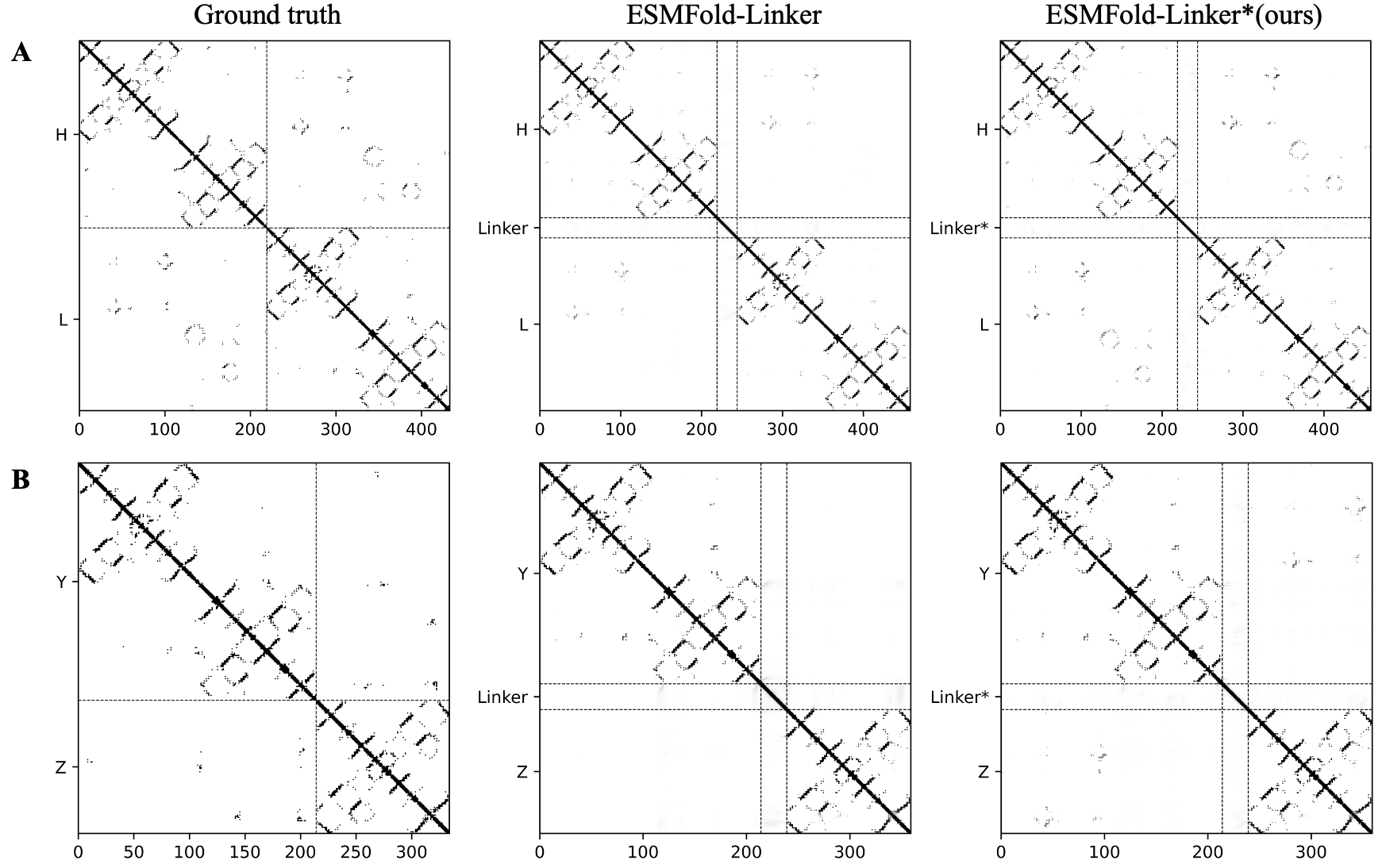}
\caption{Contact maps of viral proteins \textbf{7VYR\_HL} \textbf{(A)} and \textbf{7WPE\_YZ} \textbf{(B)}.}
\label{fig:contacts}
\end{figure}

\paragraph{The learned linker allows more chain twist while rarely interacting with the chains}
% Any interesting findings about the linker?
In Figure \ref{fig:contacts}, we visualize the predicted contact maps of two proteins with the linker inside to understand how the linker interacts with the chains. The two proteins are 7VYR\_HL and 7WPE\_YZ, corresponding to a good case (0.77 DockQ score) and a bad case (0.01 DockQ score) in our model ESMFold-Linker*. 
%In the case of 7VYR\_HL, our model achieves a 0.77 DockQ score and a 0.95 TM-score while ESMFold-Linker achieves a 0.17 DockQ score and a 0.64 TM-score. 
As shown in Figure \ref{fig:contacts}, both the G25 linker (middle) and our learned linker (right) seem to rarely interact with the protein chains in both cases. This result indicates that ESMFold is able to recognize the linker part as a disordered region and fold the connected sequences as multi-domain proteins. Furthermore, there are more predicted contacts using the learned linker than using the G25 linker in both cases. This result suggests that the learned linker allows the connecting chains to freely twist and rotate to recruit binding partners more than the manual linker. 
%We have trained two models ESMFold-Linker* and ESMFold-Linker*-Gap and thus obtain two optimized linkers in the embedding space. We project the embedding to the 20 amino acids and obtain the nearest discrete linkers: \texttt{GGGGGGGGGGGMGGGGGGGGGGGGG} and \texttt{MGRGYYYESRSGSGGGGGGGGGIIH}.

\paragraph{Limitations}
The limitations of our method include: 1) If the base model (ESMFold-v1) is not good at predicting a certain type of protein complexes, such as the heterodimers in HeteroTest2, adding an optimized linker can not make it a strong model for that type of data since the trainable parameter size is very small. 2) Our method is tested on heterodimers, whether it generalizes to homodimers or multi-chain proteins needs to be further tested. 
3) The linker is only optimized at the Folding Module, while the linker at ESM2 remains constant, which could be further improved. 4) The linker length is treated as a hyperparameter, which could be optimized to improve performance and speed.

\section{Conclusions}
% The use of prompts in protein structure prediction models is not always clear due to the model`'s high complexity and a general lack of biological knowledge for AI researchers.
% In this work, we propose Linker-tuning, a prompt tuning method designed to adapt the single-chain pre-trained ESMFold for heterodimer structure prediction. 
% As a proof-of-concept, we showcase that we can place a soft prompt in the middle of the input sequence in Folding Module of ESMFold. Under the biological prior, the complex structure prediction task is reformulated as a pre-trained task itself. 
% Our findings suggest that merely tuning a prompt can significantly improve the predicted complex structure quality over the discrete prompt handcrafted with strong biological insight. 
% Furthermore, we find that the learned linker generalizes well to OOD data and antibody data. Since our model is both efficient and fast, we think it would be promising to use it in antibody design. Moreover, if we map the learned linker to amino acids, linker optimization is in fact a sequence design problem. Hence, it might be possible to design protein sequences using our Linker-tuning framework. In future work, we plan to explore structural-based protein design based on our method. We also believe that more work can be done to further improve the performance of protein complex structure prediction by exploiting parameter-efficient tuning techniques from the NLP community.

% Connecting protein science with NLP
The use of prompts in protein structure prediction models is not always clear due to the  high complexity of models and a general lack of biological knowledge for AI researchers. In this work, we have proposed Linker-tuning, a prompt tuning method to adapt the single-chain pre-trained ESMFold for heterodimer structure prediction. As proof-of-concept, we showcase that we can place a soft prompt in ESMFold. The task is reformulated as a pre-trained task itself under the biological prior. Experiments show that merely tuning a prompt on ESMFold can significantly improve the predicted complex structure quality over the discrete prompt handcrafted with strong biological insight. We believe that better fine-tuning single-chain models holds great promise in enhancing the performance of multimer structure prediction.

\section*{Acknowledgements}

\newpage

\medskip
{
\small
\bibliography{main}
\bibliographystyle{unsrt}  %plain
}

\newpage

\appendix

\section*{Appendix}

\section{Preliminary analysis of different backbones on the inter-chain contact prediction task}
\label{preliminary}

%Preliminary analysis on inter-chain contact prediction  shows that using Folding Module on top of ESM2-3B increases prediction precision dramatically over ESM2-3B while ESM2-3B just performs slightly better ESM2-650M, implying that Folding Module is more sensitive to structure prediction and easier to control. 

We perform a preliminary experiment on the heterodimer test set on the inter-chain contact prediction task to investigate how different backbones affect performance. We compare three backbones, including ESM2-650M\footnote{https://dl.fbaipublicfiles.com/fair-esm/models/esm2\_t33\_650M\_UR50D.pt}, ESM2-3B\footnote{https://dl.fbaipublicfiles.com/fair-esm/models/esm2\_t36\_3B\_UR50D.pt}, and ESMFold-v1 which contains 3B+690M parameters.

For ESM2-650M and ESM2-3B, we train a distogram head on top of each model using single chains and their distograms from the heterodimer training set. We leverage the same architecture for the distogram head as in LM-design \cite{verkuil2022language}. 
The distogram head contains only a linear layer with a softmax activation function. Given a protein sequence, it takes the stacked attention map from ESM2 as input and outputs a probability distribution on 18 distance bins for each residue pair. Specifically, the stacked attention map for each protein is a matrix of shape (num\_layers*num\_heads, $N$, $N$), where num\_layers denotes the number of layers in ESM2, num\_heads denotes the number of attention heads in ESM2, and $N$ denotes the number of residues in the protein. For ESM2-650M, num\_layers is 33,  and num\_heads is 20. For ESM2-3B, num\_layers is 36, and num\_heads is 40. We train the distogram head with ESM2 frozen for 10 epochs using the distogram loss and select the best checkpoint based on the validation distogram loss. For ESMFold, we use its pre-trained distogram head.

% \begin{figure}[h]
% \centering
% \includegraphics[width=1\columnwidth]{images/distogram_head.png}
% \caption{Distogram head}
% \label{fig:distogram_head}
% \end{figure}

\begin{figure}[h]
\centering
\includegraphics[width=0.6\columnwidth]{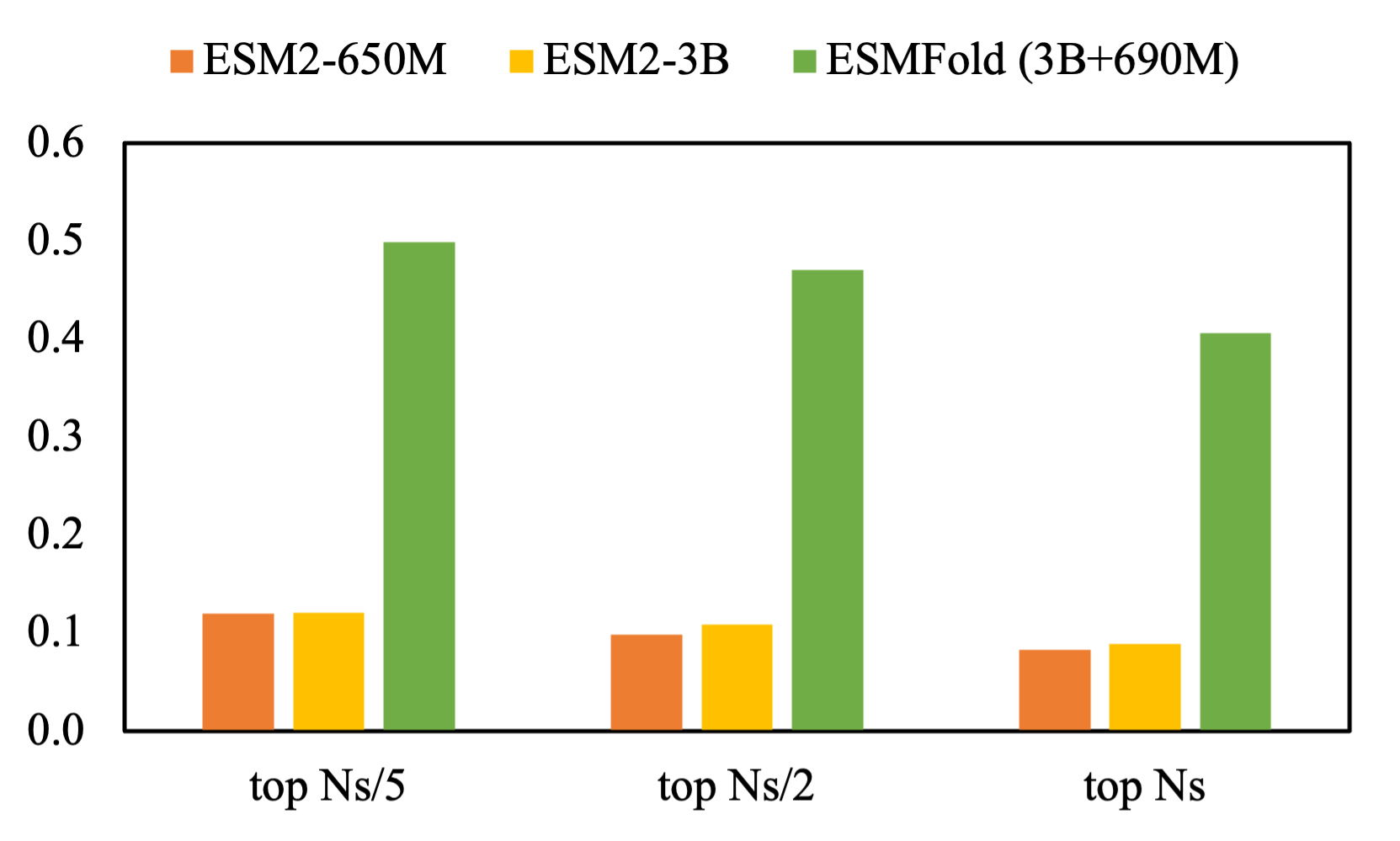}
\caption{Inter-chain contact prediction results on the heterodimer test set using different backbones, including ESM2-650M, ESM2-3B, and ESMFold (3B+690M). Chains in the heterodimer are joined by the G25 linker before input into the backbone model. }
\label{fig:esm}
\end{figure}

Figure \ref{fig:esm} shows the inter-chain contact prediction precisions on the heterodimer test set using different backbones. 
ESM2-3B performs only slightly better than the ESM2-650M while containing >2B more parameters. In contrast, ESMFold significantly improves performance over ESM2-3B while containing only 690M more parameters.
This result indicates that the Folding Module plays a critical role in extracting the structure information out of the sequence. Therefore, we suspect that ESM2 is not so sensitive to structure prediction and is harder to control. Although it is natural to place the prompt at ESM2 for NLP researchers, however, for protein structure prediction, Folding Module is a better choice for prompt tuning.

\section{Effect of linker length}
\begin{figure}[t]
\centering
\includegraphics[width=1\columnwidth]{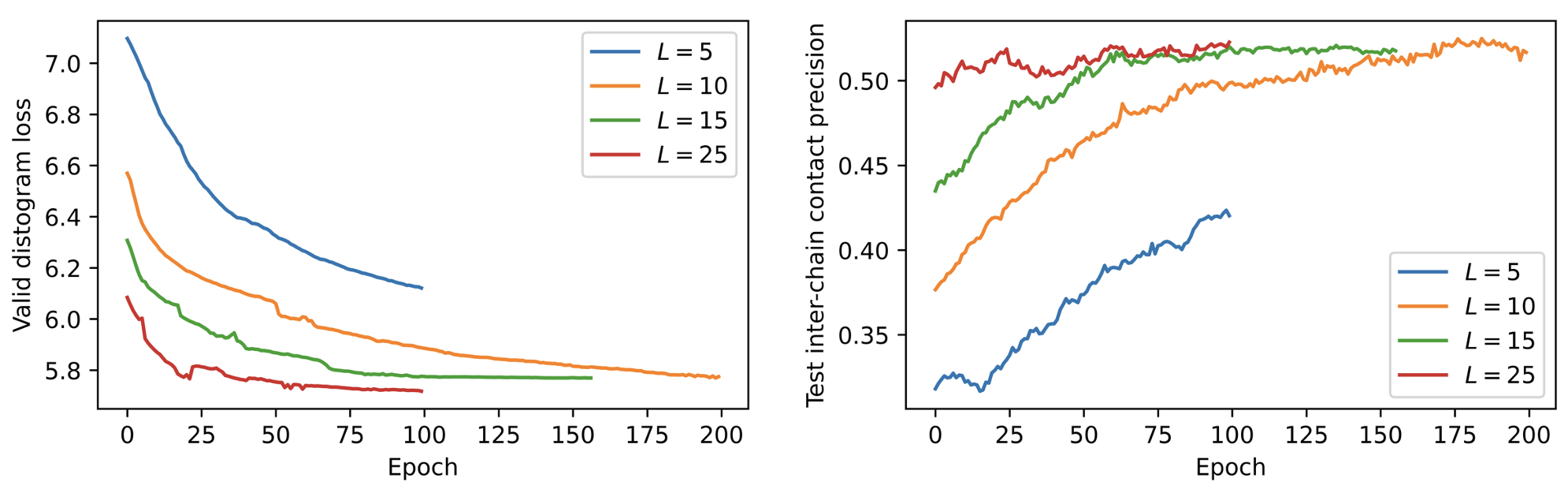}
\caption{Effect of linker length $L$.}
\label{fig:length}
\end{figure}

We study how the linker length $L$ affects the performance of our model. For each linker length $L$ in \{5, 10, 15, 25\}, we train a corresponding model on the heterodimer training samples with sequence lengths less than or equal to $220-L$. The number of training data for each model is less than 614. 
For all models, the number of recycles is set to 1 and residue\_index\_offset is set to 0 during both training and inference. 
Figure \ref{fig:length} shows the weighted distogram loss on the validation set (left) as well as the top Ns/5 inter-chain contact precision on the heterodimer test set (right) with different linker lengths. As shown in Figure \ref{fig:length}, we can see that when $L=5$, the inter-chain contact performance of the model after training 100 epochs is much worse than the performance of the other three models with $L\geq10$. 
The performance improves as $L$ increases, and they achieve comparable performance when training to convergence. However, for models with smaller $L$, it takes longer to converge during training. 
In general, $L$ should not be too small ($\leq5$). We can choose the value of $L$ in the range of $L\geq10$, depending on the training and inference speed trade-off. Empirically, the model with $L=25$ enjoys a fast convergence in training, although the inference time increases slightly (+0.8 seconds as shown in Table \ref{tab:time}) than not using the linker.

\section{Effect of weighted distogram loss}

We study how the value of $\lambda$ in the weighted distogram loss affects the performance of our model. For each $\lambda$ in \{2, 4, 6\}, we train a corresponding model for 20K steps. For all models, the number of recycles is set to 1 and residue\_index\_offset is set to 0 during both training and inference. 
The top Ns/5 inter-chain contact precisions for models with $\lambda=2,4,6$ on the heterodimer test set are 47.04\%, 47.86\%, and 47.57\%, respectively. Although the model with $\lambda=4$ achieves the best result, there is not much difference between the performances of the three models. It indicates that our method is not sensitive to the value of $\lambda$. Empirically, we set $\lambda=4$.
 
Furthermore, for the model with $\lambda=4$, we record the validation weighted distogram loss, validation inter-chain contact precision, and test inter-chain contact precision for every epoch during training. We find a Pearson correlation coefficient of -0.75 between the test inter-chain contact precision and the validation weighted distogram loss, which is larger in absolute value than the Pearson correlation coefficient of 0.47 between the test and validation inter-chain contact precision. This result suggests that the weighted distogram loss on the validation set correlates well with the quality of the predicted interface. Therefore, we recommend using the weighted distogram loss rather than the inter-chain contact precision as the model selection metric.

% \section{Metric distribution on heterodimer data}

% \begin{figure}[t]
% \centering
% \includegraphics[width=1\columnwidth]{images/hetero_boxplot.png}
% \caption{Boxplots of RMSD score and DockQ score of Heterodimer test set (\textbf{A}) and HeteroTest2 (\textbf{B}), respectively.}
% \end{figure}

% \section{Data availability}
% The training, validation, and test sets are available in the folder named data.

% \section{Code availability}
% The codes for Linker-tuning are available in the folder named code. 
% The model checkpoints are available in the folder named checkpoint.

%\section{Effect of number of training samples}
% 分析需要多少数据来训练linker 

%%%%%%%%%%%%%%%%%%%%%%%%%%%%%%%%%%%%%%%%%%%%%%%%%%%%%%%%%%%%

\end{document}